\documentclass{emulateapj}

\pdfoutput=1

\usepackage{epsfig}
\usepackage{epstopdf}
\usepackage{wrapfig}
\usepackage{graphicx}
\usepackage{rotating}
\usepackage{color}
\usepackage{subfigure}

\RequirePackage{natbib}

\newcommand{\subscript}[1]{_{\mathrm{#1}}}
\newcommand{\superscript}[1]{^{\mathrm{#1}}}

\shorttitle{Reverse Shock Dust Destruction}
\shortauthors{Silvia, Smith, \& Shull}
\slugcomment{Accepted by the Astrophysical Journal}

\begin{document}

\title{Numerical Simulations of Supernova Dust Destruction.\\
	II.  Metal-Enriched Ejecta Knots}
	
\author{Devin W. Silvia$\superscript{1}$, Britton D. Smith$\superscript{1, 2}$, and J. Michael Shull$\superscript{1}$}
\affil{$\superscript{1}$CASA, Department of Astrophysical and Planetary Sciences, University of Colorado, UCB 389, Boulder, CO 80309, USA; devin.silvia@colorado.edu; michael.shull@colorado.edu \\
$\superscript{2}$Department of Physics and Astronomy, Michigan State University, East Lansing, MI 48824, USA; smit1685@msu.edu}

\begin{abstract}
Following our previous work, we investigate through hydrodynamic simulations the destruction of newly-formed dust grains by sputtering in the reverse shocks of supernova remnants.  Using an idealized setup of a planar shock impacting a dense, spherical clump, we implant a population of Lagrangian particles into the clump to represent a distribution of dust grains in size and composition. We vary the relative velocity between the reverse shock and ejecta clump to explore the effects of shock-heating and cloud compression.  Because supernova ejecta will be metal-enriched, we consider gas metallicities from $Z/Z\subscript{\odot} = 1$ to $100$  and their influence on cooling properties of the cloud and the thermal sputtering rates of embedded dust grains.  We post-process the simulation output to calculate grain sputtering for a variety of species and size distributions.  In the metallicity regime considered in this paper, the balance between increased radiative cooling and increased grain erosion depends on the impact velocity of the reverse shock.  For slow shocks ($v\subscript{shock} \le 3000$~km~s$^{-1}$), the amount of dust destruction is comparable across metallicities, or in some cases is decreased with increased metallicity.  For higher shock velocities ($v\subscript{shock} \ge 5000$~km~s$^{-1}$), an increase in metallicity from $Z/Z\subscript{\odot} = 10$ to $100$ can lead to an additional 24\% destruction of the initial dust mass.  While the total dust destruction varies widely across grain species and simulation parameters, our most extreme cases result in complete destruction for some grain species and only 44\% dust mass survival for the most robust species.  These survival rates are important in understanding how early supernovae contribute to the observed dust masses in high-redshift galaxies.
\end{abstract}


\section{Introduction}\label{intro}
Over the last two decades, far-infrared (FIR) and millimeter observations of high-redshift quasars ($z > 6$) have produced estimates for galactic dust masses as high as $10^{8}~M_{\odot}$ \citep{Smail:1997gf, Hughes:1998ve, Bertoldi:2003fr, Wang:2008rt}.  In order to explain the formation of this large quantity of dust within the short lifetime ($\sim$1 Gyr) of the universe at this epoch, a mechanism must exist which is capable of both significant and rapid dust production.  One recently pursued solution is that the majority, if not all, of this dust comes from core-collapse supernova (CCSN) explosions of the first generations of stars \citep{Morgan:2003lr, Maiolino:2004bh, Hirashita:2005vn}.

While various theoretical work \citep{Kozasa:1989ve, Kozasa:1991ul, Todini:2001qf, Nozawa:2003pd, Bianchi:2007fp, Nozawa:2010yq} has suggested that $\sim$0.1-0.3~$M_{\odot}$ of dust could be formed per supernova event, which is in rough agreement with the $\sim$0.2-1.0~$M_{\odot}$ required to explain high-redshift dust \citep{Morgan:2003lr, Dwek:2007lr, Dwek:2011fj, Gall:2011uq}. However, observational efforts focused on local supernova remnants (SNRs) often fall orders of magnitude short of these values \citep{Stanimirovic:2005rr, Williams:2006ly, Meikle:2007dq, Rho:2008qf, Rho:2009ul, Kotak:2009lr}.  More recent observations aimed at finding colder dust ($T < 40$~K) using AKARI and the Balloon-borne Large Aperature Submillimeter Telescope (BLAST) by \cite{Sibthorpe:2010fk}, as well as {\it Herschel} by \cite{Barlow:2010lr} and \cite{Matsuura:2011qy}, find larger dust masses.  \cite{Matsuura:2011qy} find $\sim$0.6~$M\subscript{\odot}$ of dust in SN1987A.  Such estimates could make the argument for SNe as dust factories more plausible.

One question that arises from these SNR studies is what fraction of the freshly formed dust predicted by theory will survive the interaction between the reverse shocks and the ejecta.  As the fast-moving dust ($V_{ej} \ge 1000$~km~s$^{-1}$) is impacted by the reverse shock, it will be subject to sputtering and grain-grain collisions as the density and temperature of the dust-enriched gas are increased.

In our previous work \citep[][hereafter Paper I]{Silvia:2010lr}, we investigated, through hydrodynamic simulations, the destruction of newly formed dust grains by sputtering in the reverse shocks of SNRs. Using ``cloud-crushing" simulations \citep{Woodward:1976pd, Mac-low:1994fj, Klein:1994uq}, we found that the degree of dust destruction depends heavily on the initial radius distribution of the dust grains as well as the initial density of the ejecta cloud and relative velocity between the reverse shock and the cloud.  In the most extreme cases, we found grain destruction to vary from 20-100\% depending on grain species.  We also found morphological similarities within these simulations to the observational studies of Cassiopeia A presented in \cite{Fesen:2011kx}, specifically in the scenarios where the effects of cooling led to the fragmentation of ejecta clouds.

However, our earlier simulations computed the radiative cooling and sputtering rates for approximately solar metallicity ($Z_{\odot}$) gas.  In reality, the gas contained in supernova ejecta will be highly metal-enriched. The question arises as to the balance between the enhanced radiative cooling and the increase in sputtering yields in this high-metallicity regime.  This balance will also be influenced by the changing density of the metal-enriched gas as the cloud is shredded and the plasma thins out.   

In the current paper, we perform additional cloud-crushing simulations to study the evolution of dust mass contained within an idealized ejecta knot as it is impacted by a supernova reverse shock.  In Section \ref{methods} we give a brief review of the code used to carry out these simulations, the methods for tracking our dust populations, and the changes made from our previous work.  In Section \ref{sims}, we describe the simulations unique to this paper, specifically aimed at probing the higher metallicity regimes expected to be present in supernova ejecta.  We present the results of these simulations in Section \ref{results} and conclude in Section \ref{summary} with a summary and discussion of the implications of these results.

\section{Methodology}\label{methods}

\subsection{Code and Simulation Setup}\label{code}

As in Paper I, we use the Eulerian adaptive mesh refinement (AMR), hydrodynamics + N-body code, \texttt{Enzo} \citep{Bryan:1997qy, Norman:1999uq, Oshea:2004book}.  We do not make use of any of the cosmological or gravity-solving components of the code owing to the idealized nature of our problem.  We follow the same cloud-crushing setup as outlined in Paper I and refer the reader to that work for a detailed description.

The user-supplied parameters required to initialize the simulation are: cloud radius, $r\subscript{cloud}$; cloud temperature,  $T\subscript{cloud}$; density of the ambient medium, $\rho\subscript{m}$; initial over-density, $\chi$, of the cloud with respect to the ambient medium ($\rho\subscript{cloud} = \chi\rho\subscript{m}$); and the velocity of the shock relative to the stationary cloud, $v\subscript{shock}$.  From these input parameters, the following values must be derived in order to completely initialize the simulation: temperature of the ambient medium, $T\subscript{m}$, post-shock density, $\rho\subscript{shock}$, and post-shock temperature, $T\subscript{shock}$.  We set $T\subscript{m}$ so that the cloud remains in pressure equilibrium, while the shock-related values are calculated using the Rankine-Hugoniot jump conditions.

During runtime, we take advantage of the AMR capabilities of \texttt{Enzo} by employing the same refinement criterion as described in Paper I, increasing resolution in areas of the simulation that contain a significant fraction of cloud material.  Cells that are initially enclosed within the cloud radius are assigned a ``cloud material" value, $\xi$, that is advected with the flow in the same way as density.  When a cell exceeds a pre-defined cloud material ``mass", $m\subscript{\xi} = \xi V\subscript{cell}$, where $V\subscript{cell}$ is the cell volume, the resolution of that cell is doubled.  A more thorough description of this refinement process can be found in Paper I.

\subsection{Dust Tracking and Post-processing}\label{dusttrack}
\begin{figure}[htp]
\centering
  		\includegraphics[width=0.5\textwidth]{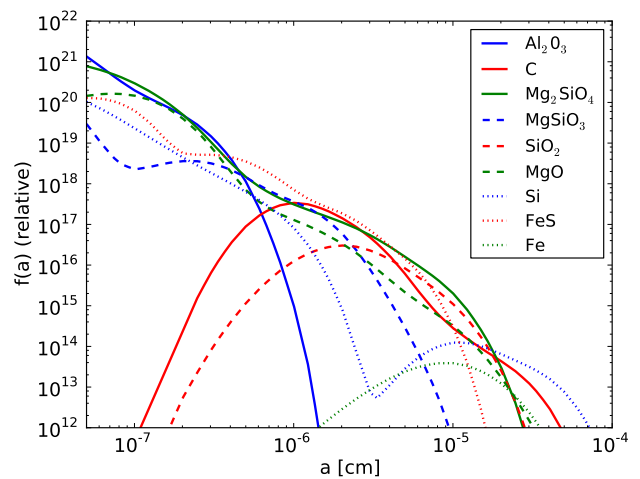}
  \caption{Reproduced grain radius distributions for the species expected in the unmixed ejecta model of a core-collapse supernova with a progenitor mass of 20 M$\subscript{\odot}$ as calculated by \cite{Nozawa:2003pd}. The $y$-axis is the abundance of grains of a given radius, $f(a)$, such that the number density of grains between $a$ and $a + \delta a$ is $n(a) = f(a) \delta a$, where $\delta a$ is set by the number of bins used to track the distribution.}\label{rad_dist}
\end{figure}

Our dust-grain populations continue to be tracked through tracer particles embedded in the flowing gas.  As before, we post-process the density and temperature histories of each tracer particle to compute dust survival rates.  For the initial distributions in grain radii, we again made use of the values calculated by \cite{Nozawa:2003pd} for a CCSN with a progenitor mass of 20~$M\subscript{\odot}$ (see Figure \ref{rad_dist}).  We follow the evolution of all nine grain species included in the unmixed grain model of \cite{Nozawa:2003pd}.  To determine the evolution in dust mass, we use the same mass proxy outlined in Paper I, which involves tracking the changes in $n(a) \times a^{3}$, where $n(a)$ is the number density of grains and $a$ is the grain radius.  However, the erosion rates used to sputter the grains with each successive time step are different than those used in Paper~I, as described in the following section.  We note that we only account for dust grain destruction in this work and do not include a prescription for possible grain growth that might occur at high densities and low temperatures.  For the majority of our simulations, the amount of time that the dust grains spend in an environment conducive to grain growth is relatively brief and any potential increase in dust mass should be minimal.

\subsection{Modifications and Additions}\label{modifications}

To account for the fact that we expect the supernova ejecta clumps to be metal-enriched, we implement both new cooling curves and grain erosion rates for high-metallicity gas.  We calculate the cooling rates using \texttt{Cloudy} \citep{Ferland:1998sf} for gas assumed to be in ionization equilibrium.  In addition, these rates assume that both the electrons and the different ion species have the same temperature.  Cooling due to the thermal emission of dust grains is not included.

\begin{figure}[htp]
\centering
	\subfigure[Metallicity-dependent cooling curves]{
  		\includegraphics[width=0.465\textwidth]{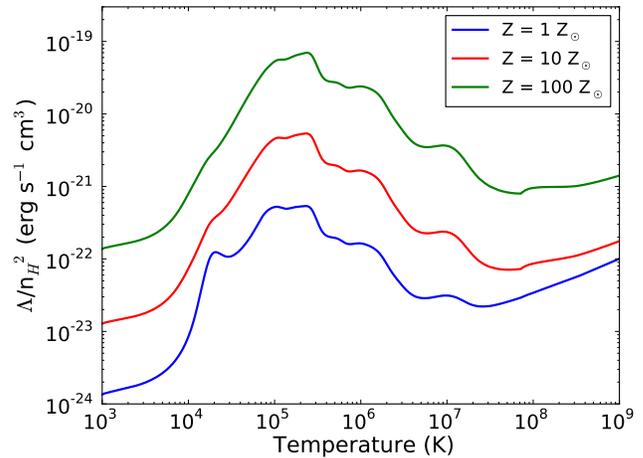}
	}
  	\subfigure[Metallicity-dependent erosion rates]{
		\includegraphics[width=0.465\textwidth]{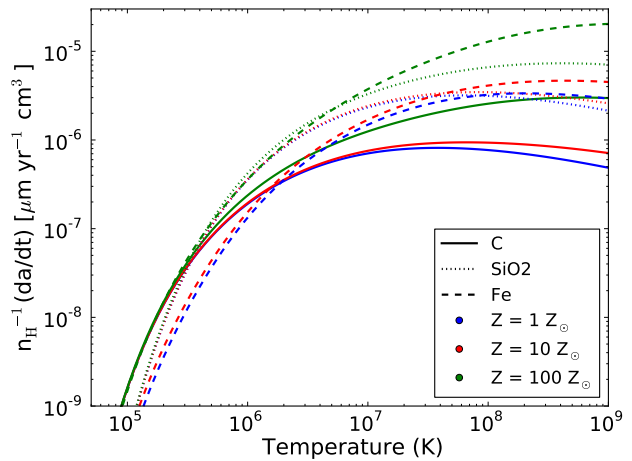}
	}
  \caption{(a) Radiative cooling curves for solar-scaled metal abundances $Z = 1~Z\subscript{\odot}$, 10~$Z\subscript{\odot}$, and 100~$Z\subscript{\odot}$. (b) Dust grain erosion rates for C, SiO$\subscript{2}$, and Fe grains.}\label{cool_erode}
\end{figure}

For this work, we compute rates for metallicities of $Z/Z\subscript{\odot} =$ 1, 10, and 100.  The element abundances for solar metallicity gas come from the composition listed in the documentation for \texttt{Cloudy}, where all abundances are specified by number relative to hydrogen.  We define the metallicities with values greater than unity to mean that the abundances for metals are increased by factors of 10 and 100 from their solar values.  The abundance of hydrogen and helium remain the same for all three metallicities.  To simplify the notation, we refer to these metallicities as 1~$Z\subscript{\odot}$, 10~$Z\subscript{\odot}$, and 100~$Z\subscript{\odot}$ for the remainder of the paper.  We also note that the values for the metal abundances remain static for the duration of our simulations, and any potential increase in metal abundance due to sputtered dust grains is omitted.  We comment on this omission in the final section of the paper.  Figure \ref{cool_erode}a shows the cooling rate coefficients for 1, 10, and 100~$Z\subscript{\odot}$ gas as a function of temperature.  To understand the contributions that hydrogen, helium, and heavier elements make on the cooling curves, we refer the reader to \cite{Gnat:2007fk}.

In contrast to Paper I, in which the cooling rates were applied to all cells within the simulation domain, we only cool those cells that contain cloud material.  This reduces the computational time required to calculate the cooling rates and accounts for the fact that, while the cloud itself can be highly metal-enriched, the ambient medium is at much lower densities and metallicities.  To ensure that important information is not lost by cooling only the cloud, we consider that for an ambient medium with a metallicity of $Z = 1~Z\subscript{\odot}$, a number density of $n = 1$~cm$^{-3}$, and a temperature of $T = 10^6$~K, the cooling time is a few~$\times~10^4$~yrs, much longer than the time scale for any of the simulations included in this work.  

We calculate new erosion rates using the formula provided by \cite{Nozawa:2006ve} for the limit in which thermal sputtering is the dominant mechanism for grain erosion.  This is appropriate, since the tracer particles used to track the dust are embedded in the flow.  In computing these erosion rates, we use the same scaled solar metal abundance ratios that were used to produce our new cooling functions.  Computing rates based on these enhanced metal abundances is an important step.  The sputtering yield, $Y(E)$ where $E$ is the energy of the impacting ion, is strongly dependent on the atomic mass of the impactor; high-mass ions can have orders of magnitude higher yields at high energies.  The differences in erosion rates as a function of metallicity are shown for carbonaceous, silicate, and ferrous grains in Figure \ref{cool_erode}b.  The differences do not become significant until the temperature exceeds a few times $10^{6}$~K and, even then, only for the highest metallicity.  In order to simplify Figure \ref{cool_erode}b, we have omitted the other six grain species studied in this work.  However, the erosion rates for the omitted species are comparable to the three presented species.

\begin{figure}[htp]
\centering
	\subfigure[Sputtering yields]{
  		\includegraphics[width=0.465\textwidth]{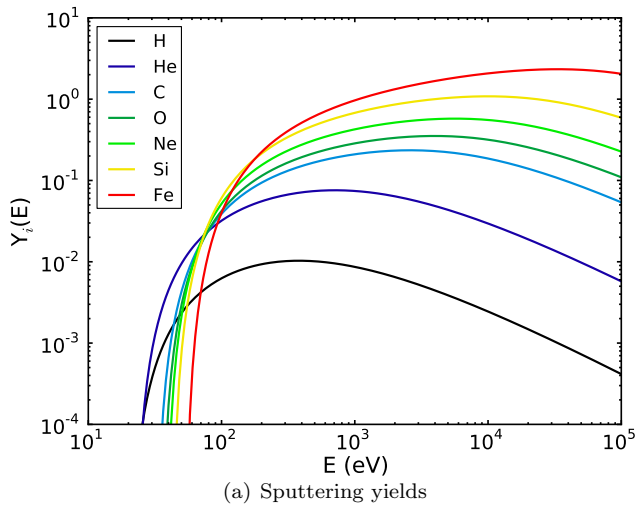}
	}
  	\subfigure[Metallicity-dependent contributions to sputtering]{
		\includegraphics[width=0.465\textwidth]{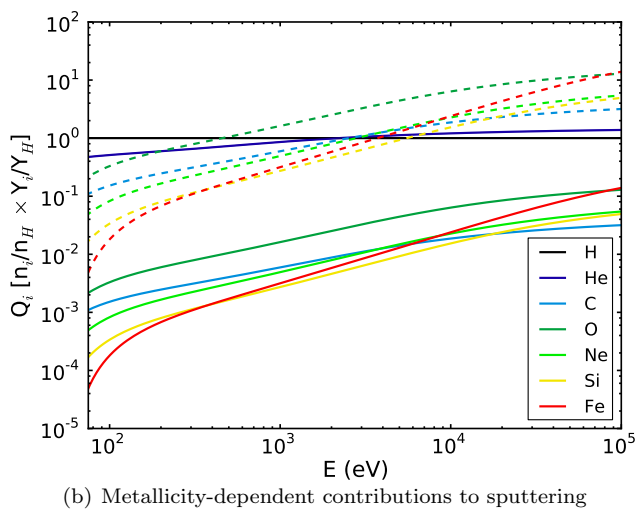}
	}
  \caption{(a) Sputtering yields, $Y\subscript{i}(E)$, of C grains versus projectile energy, E, for various incident ions; (b) The contribution of various incident ions to the sputtering of dust grains, $Q\subscript{i}$, where $Q\subscript{i}$ is defined to be the product of the element abundance by number relative to hydrogen, $n\subscript{i}/n\subscript{H}$, and the sputtering yield normalized to that of hydrogen, $Y\subscript{i}/Y\subscript{H}$, as a function of energy.  We show the contributions based on solar abundance ratios (solid lines) and abundance ratios for $Z = 100~Z\subscript{\odot}$ (dashed lines).}\label{sput_yield}
\end{figure}

We include Figure \ref{sput_yield}a to show the nature of the sputtering yields for a sample of impacting ions as a function of energy and Figure \ref{sput_yield}b to show the contribution of various impacting ions to overall grain sputtering for gas with both a metallicity of $1~Z\subscript{\odot}$ and $100~Z\subscript{\odot}$.  For simplicity, we only present these quantities for C grains, but note that the figures are qualitatively similar for the other grain species.  From Figure \ref{sput_yield}b, it is evident that the contributions from the sputtering yields of high-mass ions do not begin to dominate over hydrogen and helium until the metallicity reaches $100~Z\subscript{\odot}$, which agrees with the trend observed in the erosion rates presented in Figure \ref{cool_erode}b. 

When these new cooling and sputtering rates are implemented within our simulations, we find that, for the high metallicity ($Z = 100~Z_{\odot}$) simulations, the densest cells end up with extremely short cooling times, often shorter than the time step set by the Courant-Friedrichs-Lewy (CFL) condition.  This tends to result in cells that over-cool during a given time step and can lead to negative cell energies.  To avoid this issue, we make two modifications to \texttt{Enzo}.  First, we add a temperature floor, $T\subscript{floor} = 1000$~K, to all of our simulations to prevent clouds from becoming unrealistically cold.  Given the energetic environment of the supernova remnant and the background radiation from shock-heated, x-ray emitting gas in the remnant's shell, such a floor is physically reasonable.  Second, we modify the time-step calculation such that it is never more than 25\% of the cooling time.  While this prevents negative cell energies, it can become computationally expensive, as the cloud is compressed to high densities and the cooling time becomes very short in some regions of the cloud.  It is this computational limitation that prevents us from exploring even higher metallicities, at least within the bounds of our computational resources.

\section{Simulations}\label{sims}
All of the simulations presented in this work have root-grid dimensions of 512~$\times$~256~$\times$~256 and a physical resolution of 1.25~$\times~10^{15}$~cm ($\sim$84~AU) per cell edge.  As in our previous work, we initialize each cloud to have a radius of $r\subscript{cloud}=10^{16}$~cm, which spans 8 cells at the root-grid resolution.  We also use the same Gaussian envelope formulation as in Paper I, with the density fall-off occurring at $r = 0.7~r\subscript{cloud}$.

We allow the simulation domain to be refined up to three additional levels such that the highest resolution cells will be 8 times smaller than the root-grid.  This differs from our previous work, where we only allowed for two additional levels of resolution.  The third level of refinement was required specifically in the simulations with high metallicity, to follow the collapse of some fragments to much higher densities and smaller spatial scales than in the low-metallicity cases.  When we allow for a fourth level of refinement, the final dust masses do not change significantly; we therefore limit ourselves to three levels to save computational resources. As mentioned above, our cell refinement is based on the amount of cloud material in a given cell. 

For this work, we primarily focus our exploration of parameter space to the relative velocity between the reverse shock and the ejecta cloud and the metallicity of the gas contained within the cloud.  Specifically, we investigate shock velocities of 10$^{3}$~km~s$^{-1}$, 3$\times$10$^{3}$~km~s$^{-1}$, 5$\times$10$^{3}$~km~s$^{-1}$, and 10$^{4}$~km~s$^{-1}$ and metallicities of 1~$Z\subscript{\odot}$, 10~$Z\subscript{\odot}$, and 100~$Z\subscript{\odot}$. For ease of reference, we provide a numbered list of all simulations in Table \ref{sim_table}.  For all simulations, the cloud constrast is $\chi = 1000$, the initial cloud temperature is set equal to our temperature floor, $T\subscript{cloud} = T\subscript{floor} = 1000$~K, and the total runtime is $4.2t\subscript{cc}$, where $t\subscript{cc} = \chi^{1/2}r\subscript{cloud}/v\subscript{shock}$ is the cloud-crushing time \citep{Klein:1994uq}.

\begin{deluxetable}{lrrrrr}
\tabletypesize{\scriptsize}
\tablecaption{Simulation Parameters \label{sim_table}}
\tablewidth{0pt}
\tablehead{\colhead{{\bf Simulation}} &
	\colhead{$\chi$} &
	\colhead{$v\subscript{shock}$} &
	\colhead{$t\subscript{cc}$} &
	\colhead{Metallicity} \\
	\colhead{} &
	\colhead{} &
	\colhead{(km s$^{-1}$)} &
	\colhead{(yrs)} &
	\colhead{($Z/Z_{\odot}$)}}
\startdata
{\bf 1}	&	1000		&	1000		&	100.2	&	1	\\
{\bf 2}	&	1000		&	1000		&	100.2	&	10	\\
{\bf 3}	&	1000		&	1000		&	100.2	&	100	\\
{\bf 4}	&	1000		&	3000		&	33.4		&	1	\\
{\bf 5}	&	1000		&	3000		&	33.4		&	10	\\
{\bf 6}	&	1000		&	3000		&	33.4		&	100	\\
{\bf 7}	&	1000		&	5000		&	20.0		&	1	\\
{\bf 8}	&	1000		&	5000		&	20.0		&	10	\\
{\bf 9}	&	1000		&	5000		&	20.0		&	100	\\
{\bf 10}	&	1000		&	10000	&	10.0		&	1	\\
{\bf 11}	&	1000		&	10000	&	10.0		&	10	\\ 
{\bf 12}	&	1000		&	10000	&	10.0		&	100
\enddata
\tablecomments{Parameter definitions: $\chi$ is the initial over-density of the cloud with respect to the ambient medium, $v\subscript{shock}$ is the relative velocity between the inflowing shock and the stationary cloud, and $t\subscript{cc}$ is the cloud-crushing time.}
\vspace{5mm}
\end{deluxetable}

\begin{figure*}[htp]
\centering
	\includegraphics[width=1.0\textwidth]{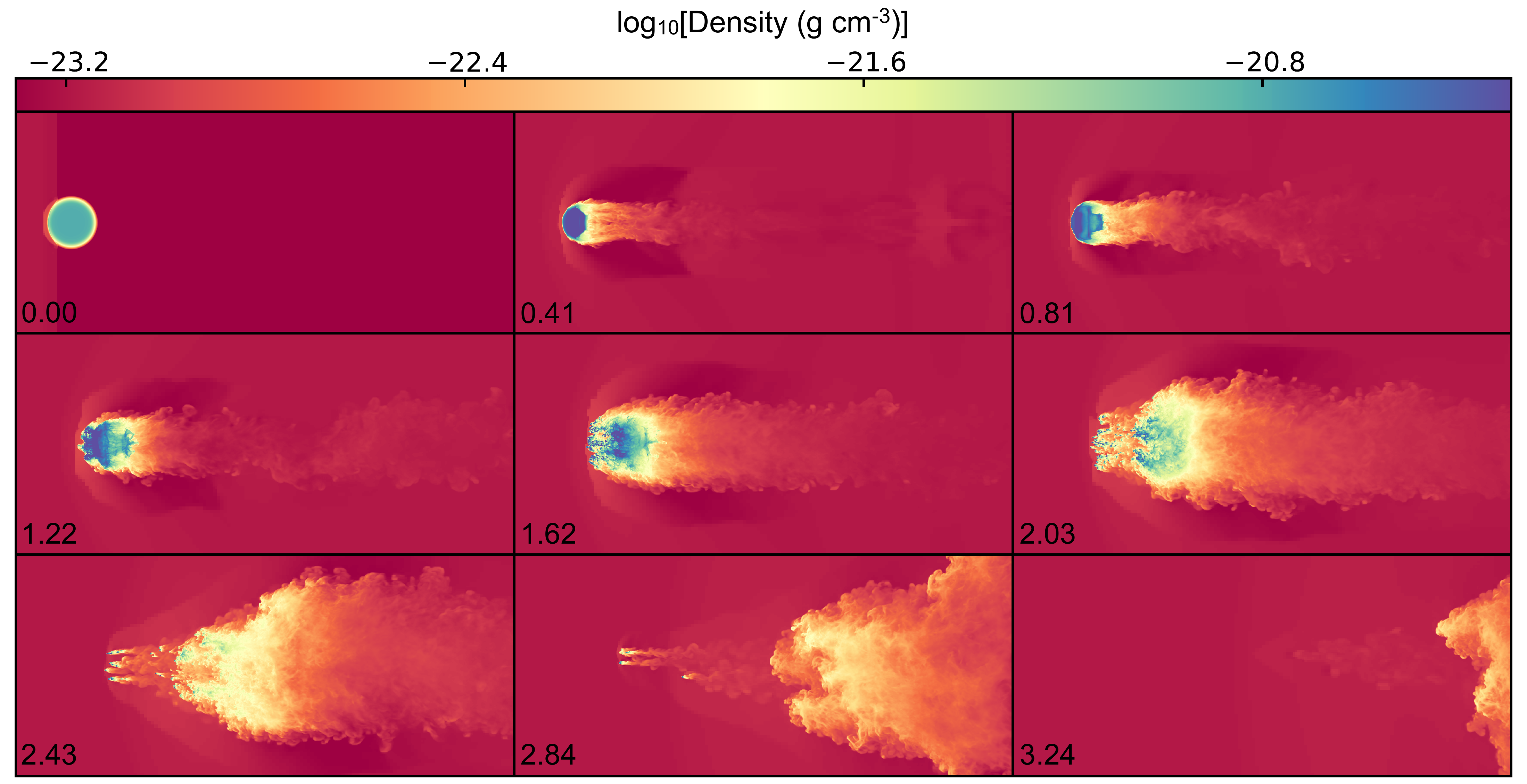}
	\caption{Density projections in the $x$-$z$ plane weighted by cloud material for the case of $v\subscript{shock} = 5000$~km~s$^{-1}$ and $Z = 100~Z\subscript{\odot}$.  The number in the lower left corner of each panel indicates the time of the snapshot in units of the cloud-crushing time, $t\subscript{cc}$.  Projections were made using the software analysis package, \texttt{yt} \citep[yt-project.org;][]{Turk:2011lr}.}\label{simulation_evol}
\end{figure*}

Figure \ref{simulation_evol} shows the evolution in density of the shock-cloud interaction from the initial impact to $t \sim 3~t\subscript{cc}$ for $v\subscript{shock} = 5 \times 10^3$ km~s$^{-1}$ and $Z = 100~Z\subscript{\odot}$.  Notable features include the tail of ablated material produced as the shock washes over the cloud, and the numerous cold, dense fragments that form a result of the high cooling rates for the metal-enriched gas.  A handful of these fragments persist for a large fraction of the total simulation time.

While our previous work ran all simulations for a total of ten cloud-crushing times, we found that in our new Simulations 2 and 3 (slow shock and high metallicities) it was difficult to keep the majority of the dust particles within the computational domain much beyond four cloud-crushing times without significantly increasing the size of the root grid.  Though possible, we deemed this solution to be a poor use of our computational resources.  As will be seen in Section \ref{results}, the dust masses are still evolving in the relatively slow-shock ($v\subscript{shock} =  10^{3}$~km~s$^{-1}$ and 3$\times$10$^{3}$~km~s$^{-1}$) simulations at the point of termination.  However, the fast-shock simulations do not have significant changes in dust mass beyond that time.  Therefore, while we do see clear trends in the slow-shock simulations as a function of metallicity, we cannot say definitively how these trends would behave at later times.

\section{Results}\label{results}
We present the results for the twelve simulations listed in Table \ref{sim_table}, with a focus on the evolution in total dust mass and the time spent by the post-processed dust particles in various areas of density-temperature phase-space.   

\subsection{Dust Mass Evolution}
\begin{figure*}[htp]
\centering
	\subfigure[Al$_2$O$_3$]{
  	\includegraphics[width=0.325\textwidth]{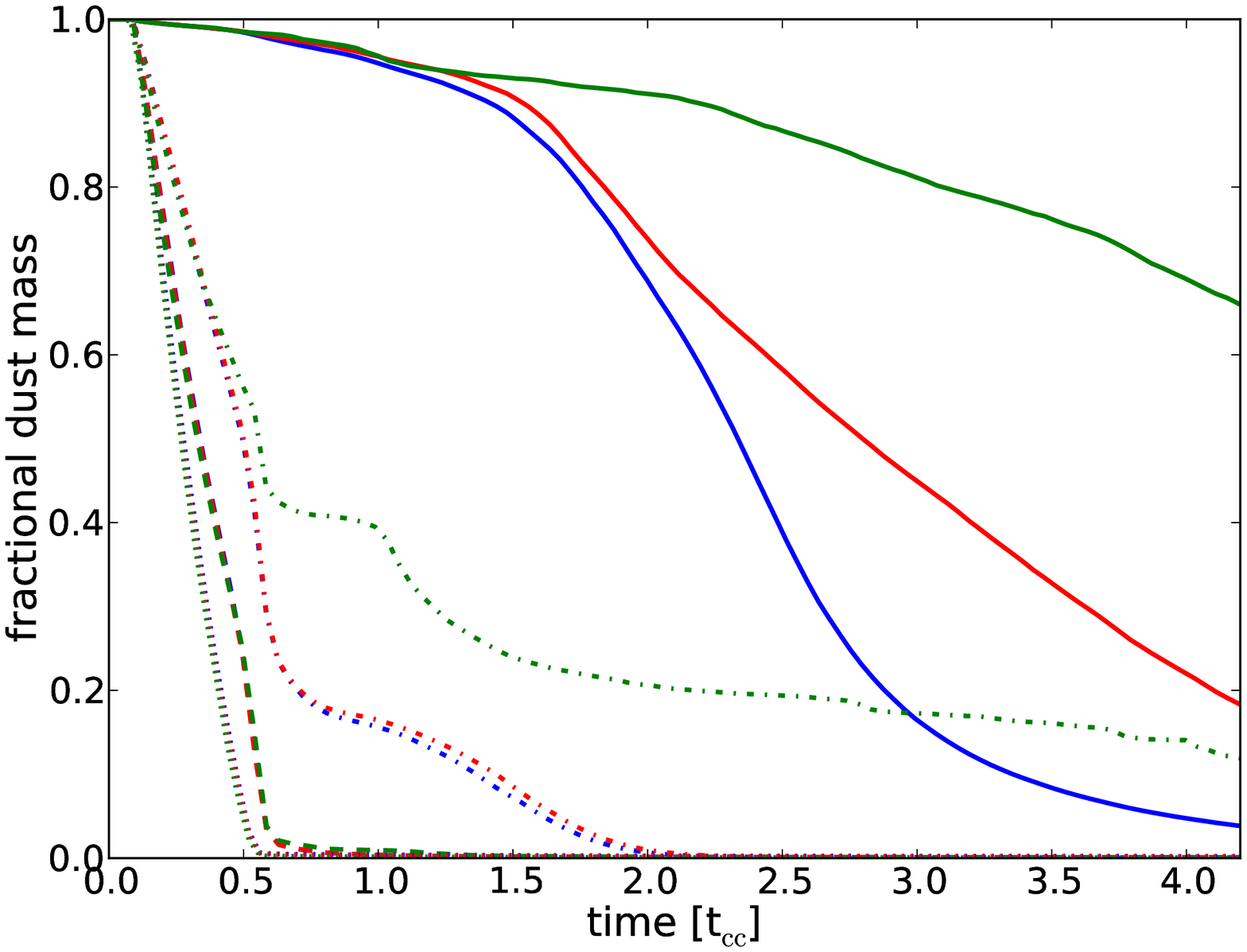}
  	}
	\hspace{-5.00mm}
  	\subfigure[C]{
  	\includegraphics[width=0.325\textwidth]{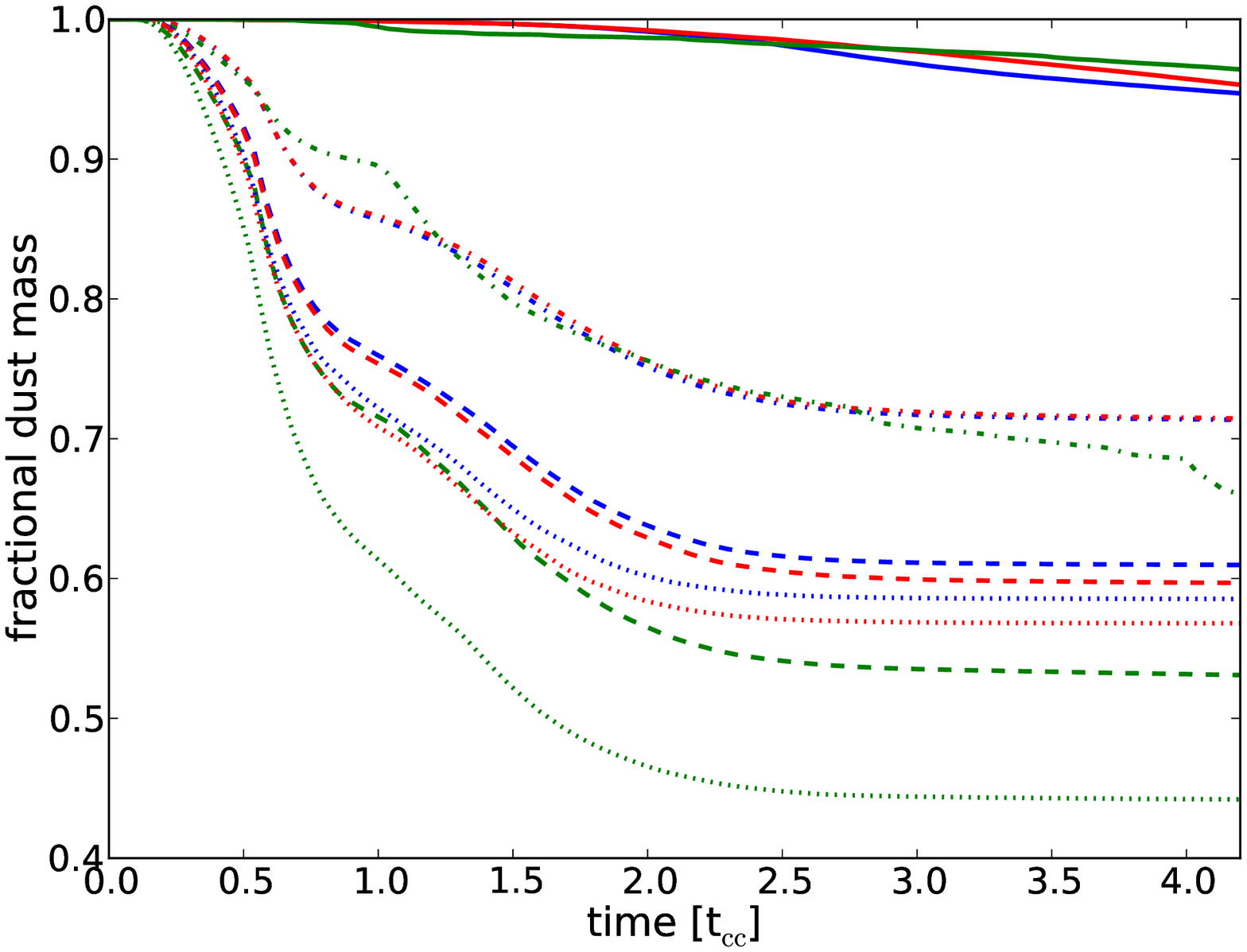}
  	}
	\hspace{-5.00mm}
	\subfigure[Fe]{
  	\includegraphics[width=0.325\textwidth]{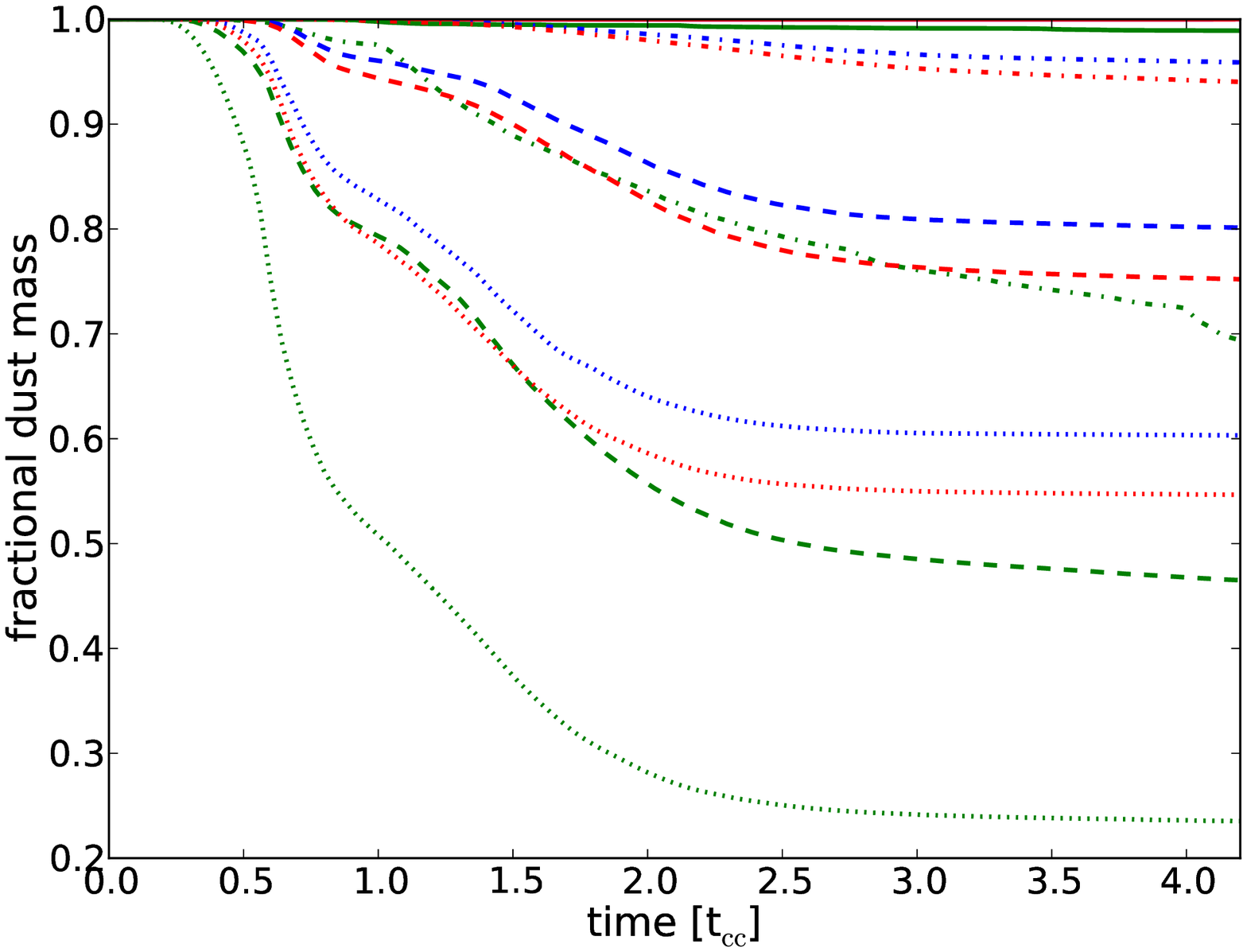}
	}
	\hspace{-5.00mm}
	\subfigure[FeS]{
  	\includegraphics[width=0.325\textwidth]{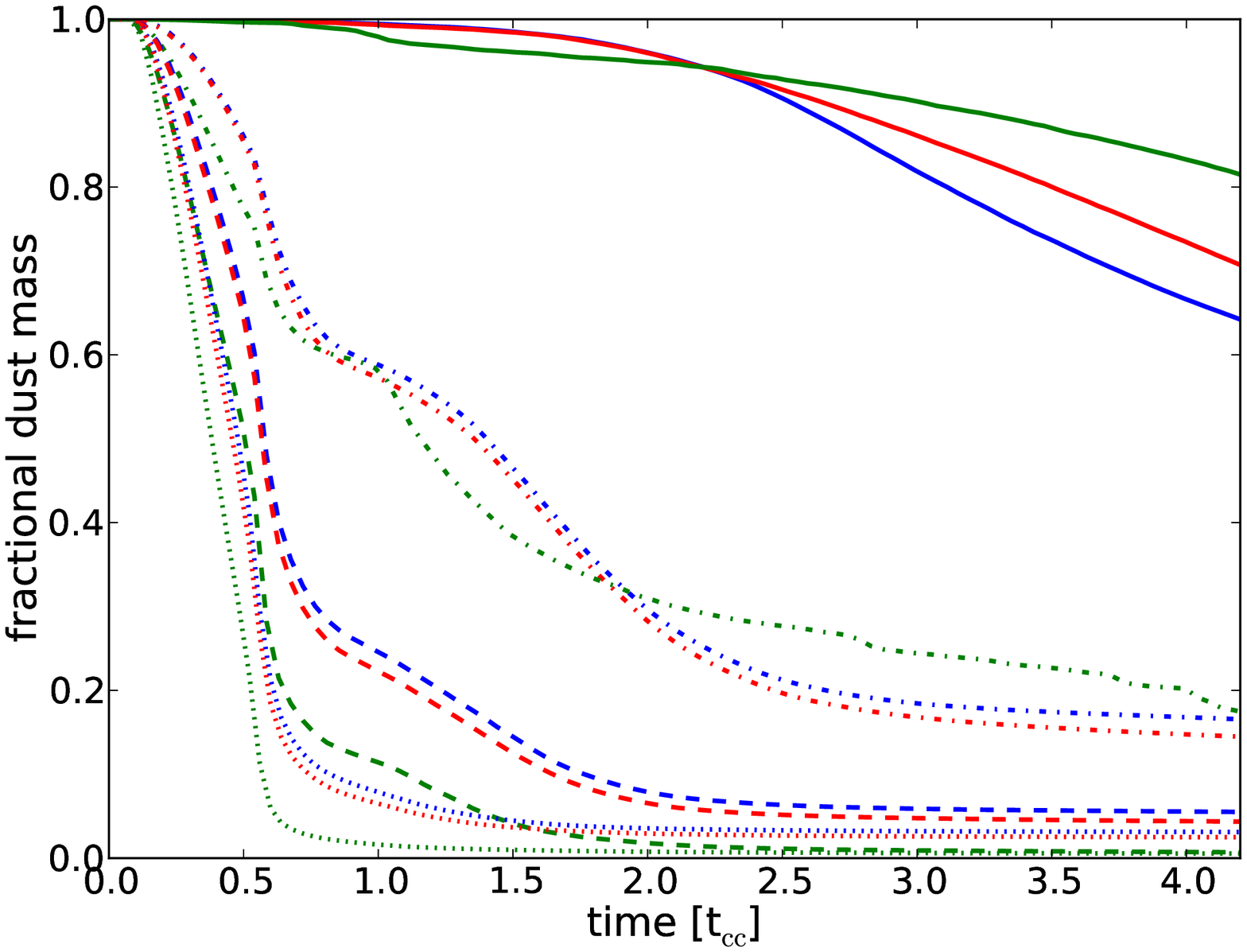}
  	}
	\hspace{-5.00mm}
	\subfigure[Mg$_2$SiO$_4$]{
  	\includegraphics[width=0.325\textwidth]{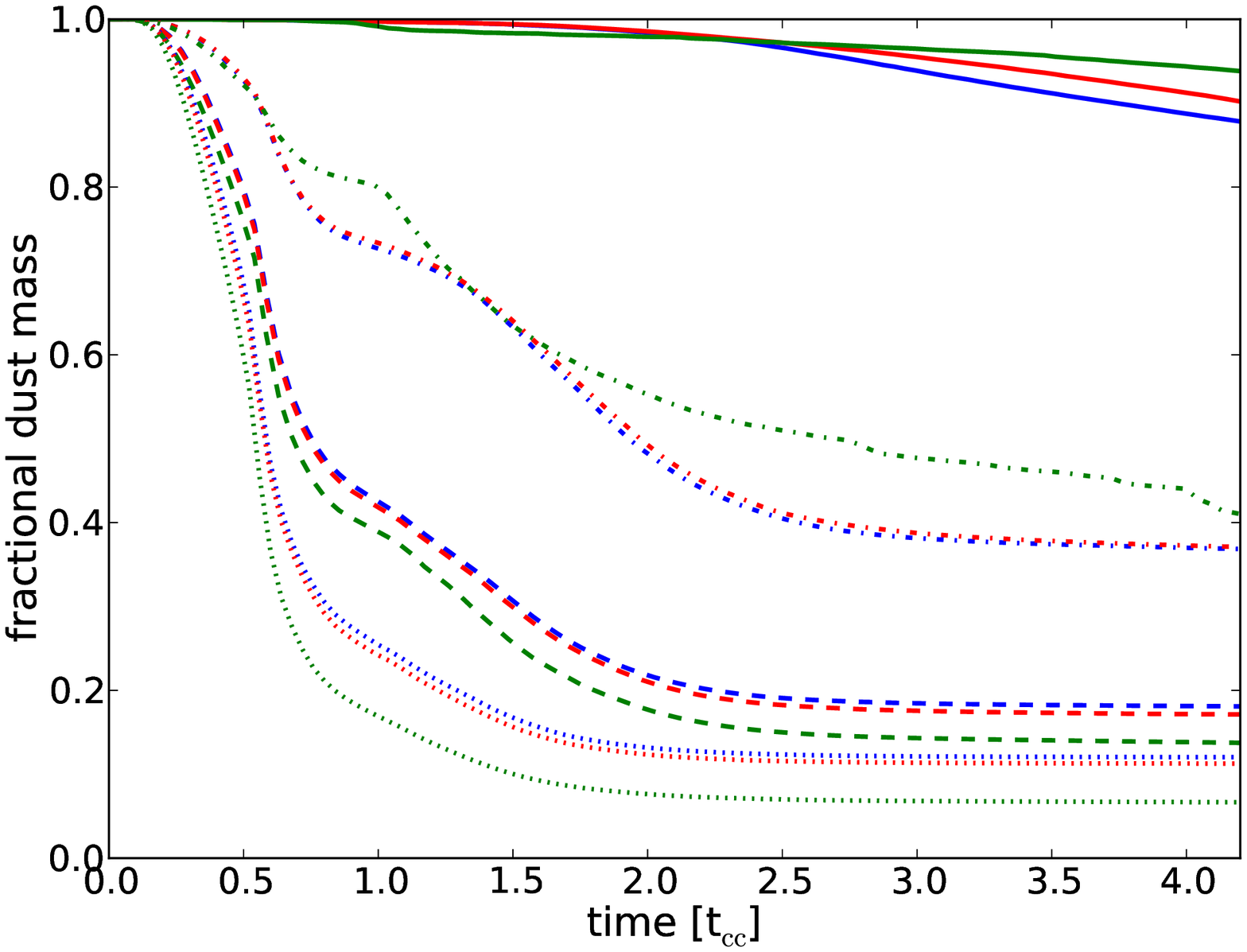}
  	}
	\hspace{-5.00mm}
	\subfigure[MgO]{
  	\includegraphics[width=0.325\textwidth]{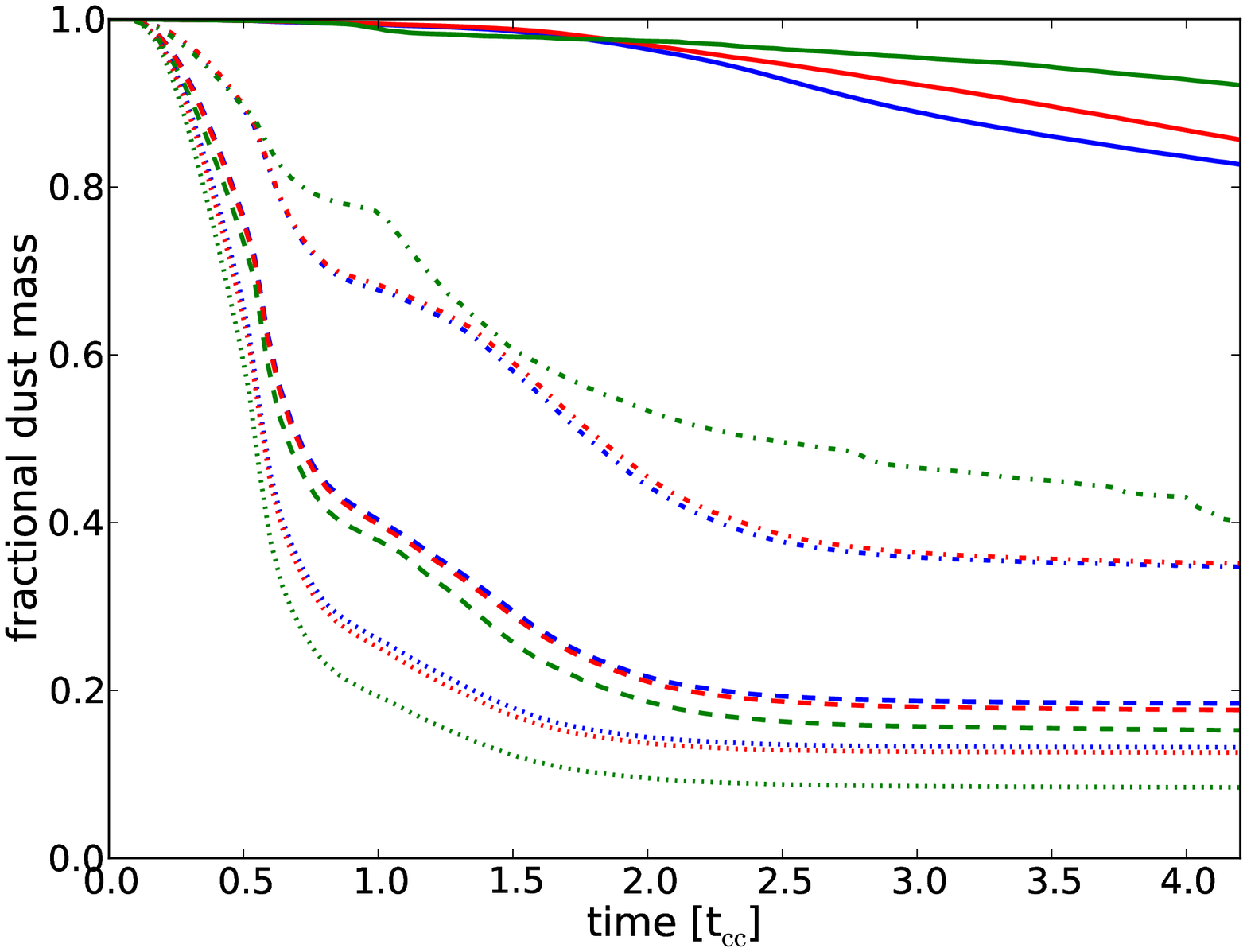}
  	}
	\hspace{-5.00mm}
	\subfigure[MgSiO$_3$]{
  	\includegraphics[width=0.325\textwidth]{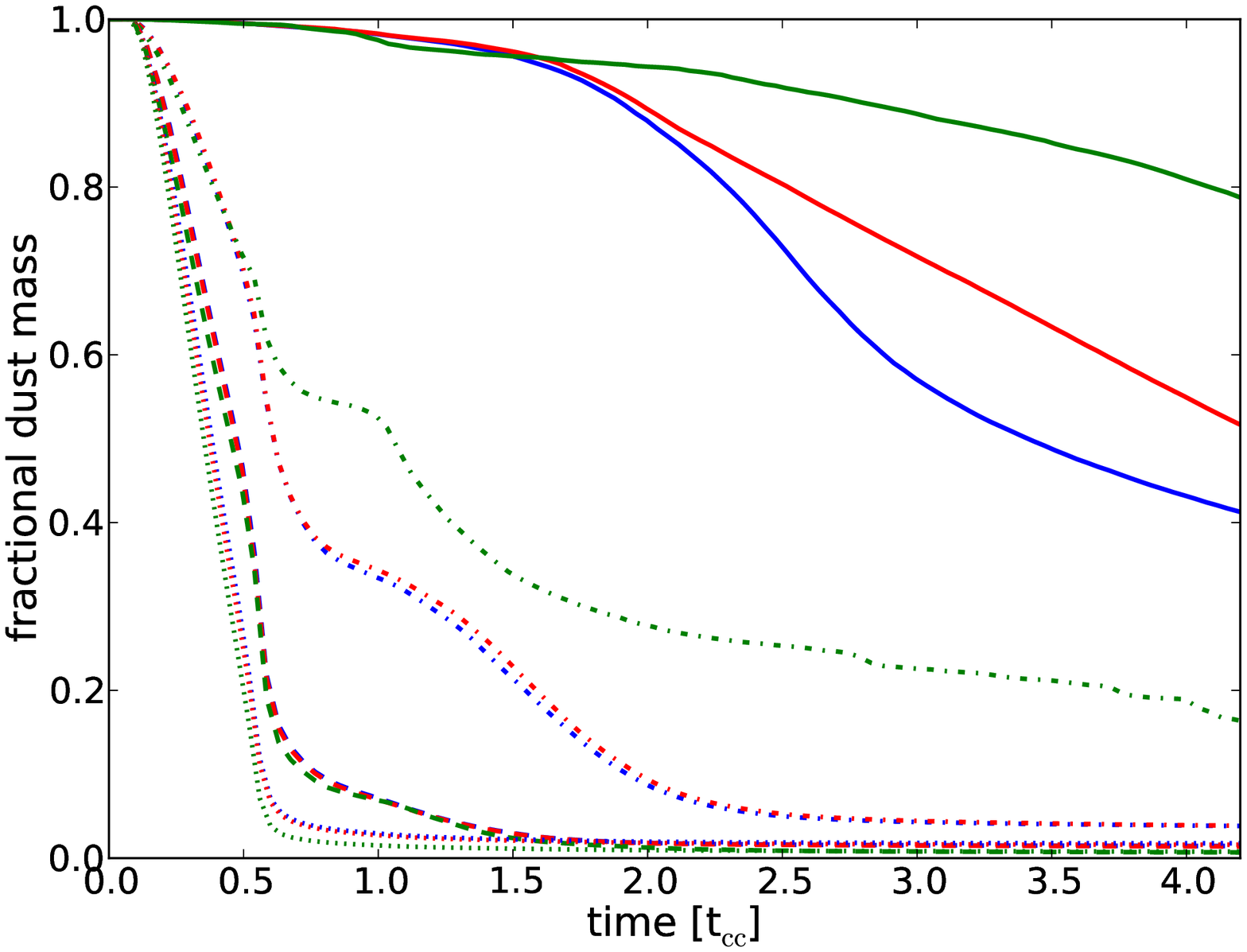}
  	}
	\hspace{-5.00mm}
	\subfigure[Si]{
  	\includegraphics[width=0.325\textwidth]{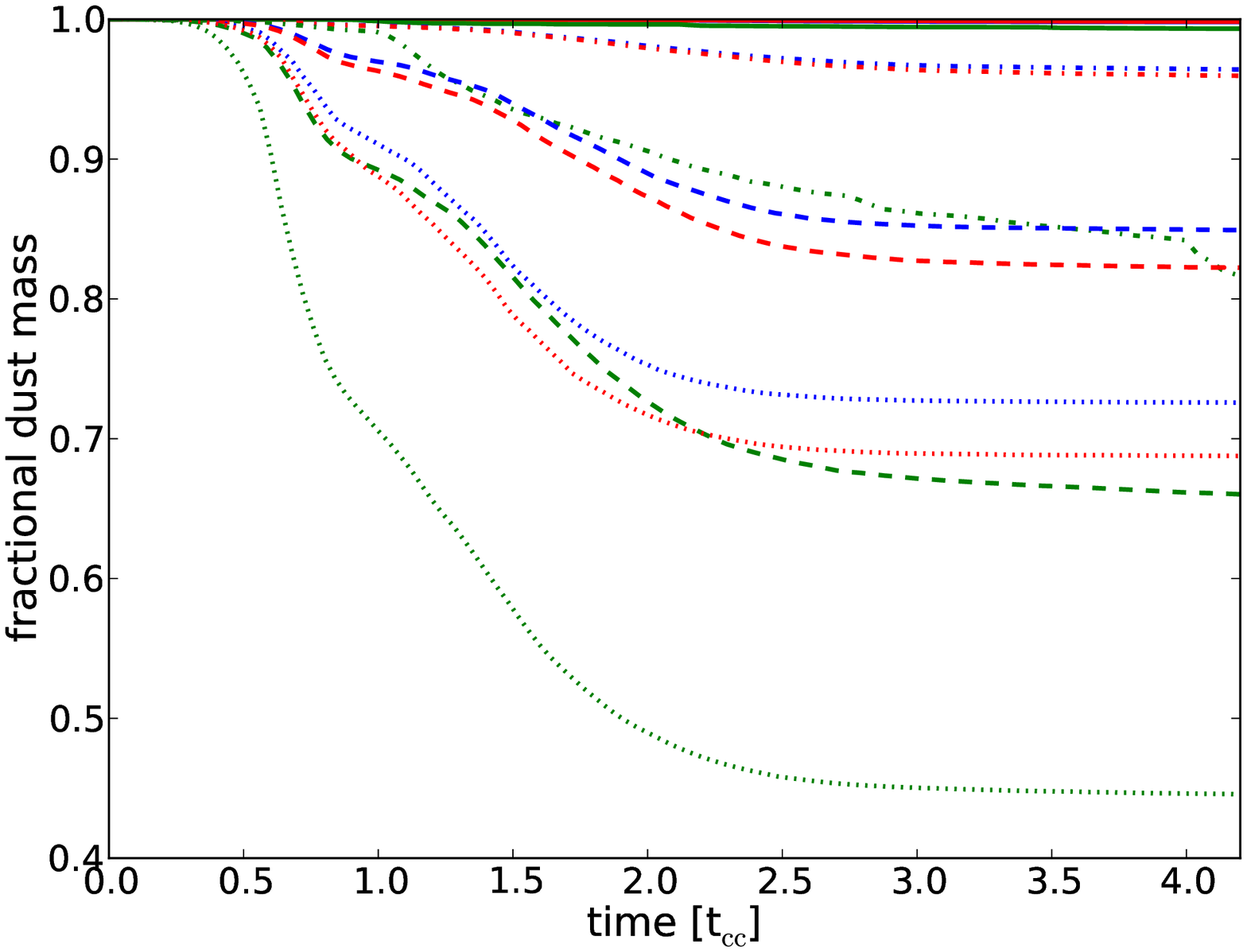}
  	}
	\hspace{-5.00mm}
	\subfigure[SiO$_2$]{
  	\includegraphics[width=0.325\textwidth]{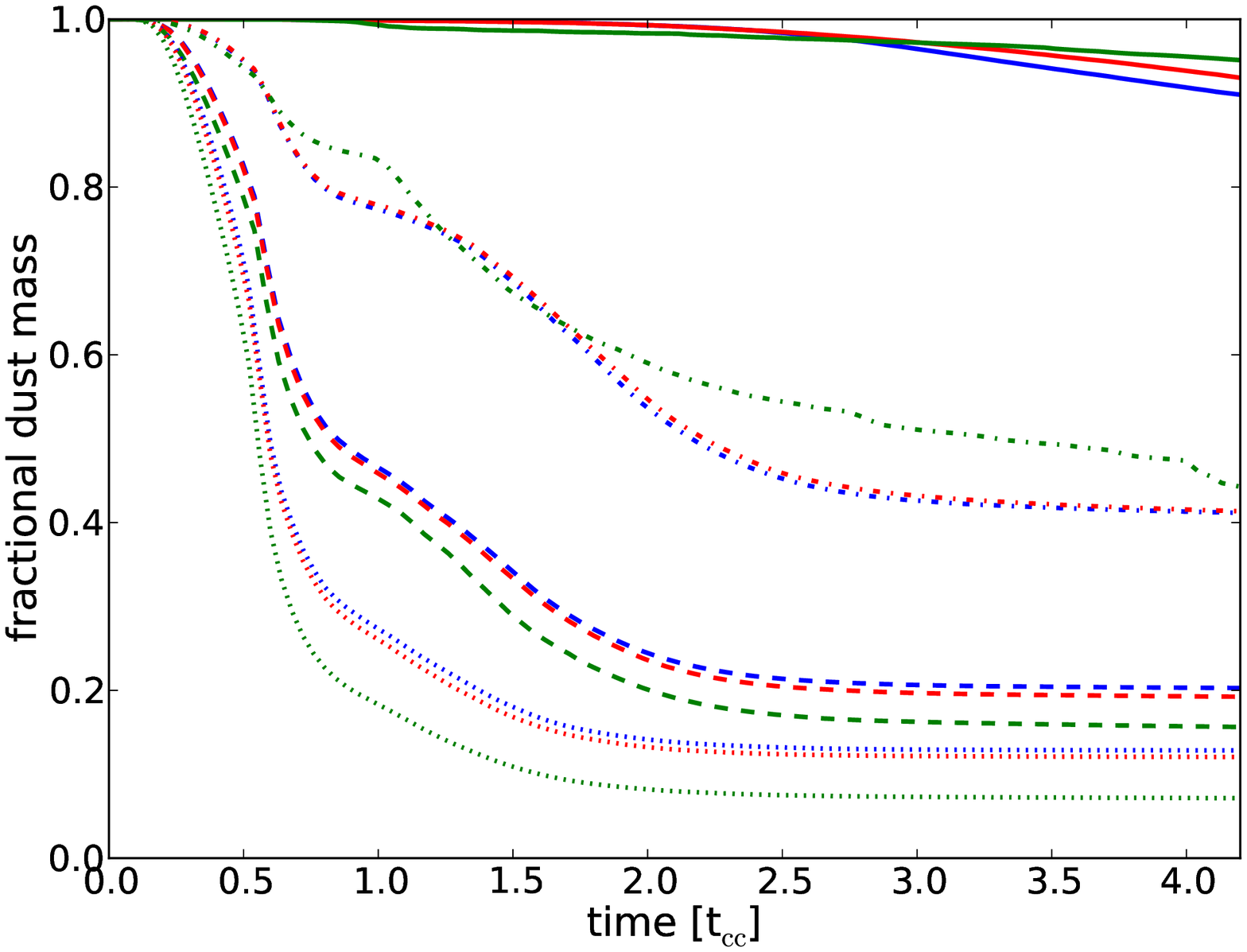}
  	}
  \caption{Dust mass evolution (surviving mass) for the nine dust species tracked in this work versus time in units of cloud-crushing time.  The line styles correspond to the four different shock velocities, $10^{3}$~km~s$^{-1}$ (solid), $3 \times 10^{3}$~km~s$^{-1}$ (dot-dashed), $5 \times 10^{3}$~km~s$^{-1}$ (dashed), and $10^{4}$~km~s$^{-1}$ (dotted).   The colors correspond to the three different metallicities, $Z = 1~Z\subscript{\odot}$ (blue), $10~Z\subscript{\odot}$ (red), $100~Z\subscript{\odot}$ (green).  Panels (b), (c), and (h) do not fully span the range 0 to 1 on the $y$-axis.}\label{dustmassevol}
\end{figure*}

The evolution of dust mass for each of the nine dust species and for all twelve simulations can be seen in Figure \ref{dustmassevol}, and the final dust masses can be located in Table \ref{dustsurvival}.  The most notable difference between the simulations is the drastic increase in dust destruction between the slowest shock velocity, $v\subscript{shock} = 10^{3}$~km~s$^{-1}$, and the highest shock velocity, $v\subscript{shock} = 10^{4}$~km~s$^{-1}$, in some cases a difference of $>$70\%.  For the intermediate shock velocities, $v\subscript{shock} = 3 \times 10^{3}$~km~s$^{-1}$ and $5 \times 10^{3}$~km~s$^{-1}$, the evolution in dust mass depends on the grain species.  For many species, the simulations with $v\subscript{shock} = 5 \times 10^{3}$~km~s$^{-1}$ show comparable dust destruction to the highest velocity shocks, while the simulations with $v\subscript{shock} = 3 \times 10^{3}$~km~s$^{-1}$ exhibit only moderate dust destruction.  There does appear to be a fundamental difference between the simulations with the slowest shock velocities and those with faster shocks at early times.  For nearly all grain species, there is a sharp drop in dust mass within one cloud-crushing time for shock velocities of $3 \times 10^{3}$~km~s$^{-1}$ and higher, with the magnitude of the drop increasing with shock velocity.  This suggests that there is a threshold velocity somewhere between $10^{3}$ and $3 \times 10^{3}$~km~s$^{-1}$ at which the initial impact of the shock into the cloud results in a degree of shock heating and compression that significantly influences the overall dust destruction in the cloud.  For slower shocks, the majority of the dust destruction occurs at much later times, as the cloud is shredded and incorporated into in the hot, post-shock gas.

In our study of the effects of metallicity for a given shock velocity, we find that only in the slowest shocks does an increase in metallicity consistently lead to a decrease in dust destruction (see the results of Simulation 3).  In the case of $v\subscript{shock} = 3 \times 10^{3}$~km~s$^{-1}$ and $Z = 100~Z\subscript{\odot}$, the amount of dust destruction is noticeably less for some grain species (Al$_2$O$_3$, MgSiO$_3$) and considerably more for others (Fe, Si).  This suggests that, for some grain species, the increased cooling due to the enhanced metallicity is able to counter-balance the shock-heating and keep the gas in a regime of decreased sputtering.  For the grain species with lower survival rates, a shock velocity of $3 \times 10^{3}$~km~s$^{-1}$ is sufficient to drive up the amount of dust destruction.  However, we again note that this is only based on the results at $t = 4.2t\subscript{cc}$.  At this velocity, the simulations with $Z = 1~Z\subscript{\odot}$ and $10~Z\subscript{\odot}$ appear to have hit a dust mass plateau, but for $Z = 100~Z\subscript{\odot}$ the evolution in dust mass still has a clear downward slope.  For $v\subscript{shock} = 5 \times 10^{3}$~km~s$^{-1}$ and $10^{4}$~km~s$^{-1}$, an increase in metallicity leads directly to an increase in total dust destruction.  The most drastic change occurs when the metallicity is increased from $Z = 10~Z\subscript{\odot}$ to $100~Z\subscript{\odot}$; $\sim$20\% more of the initial dust is lost with this change in metallicity for Fe and Si grains in the highest velocity scenario.  For all simulations in which the evolution in dust mass plateaus at late times, the cloud has been significantly shredded and the density of the gas has been reduced to a level that does not allow for appreciable grain sputtering.

\begin{deluxetable}{lccc}
\tabletypesize{\scriptsize}
\tablecolumns{10}
\tablecaption{Final Dust-mass Survival Fraction \label{dustsurvival}}
\tablewidth{0pt}
\tablehead{\colhead{{\bf Simulation}} &
	\colhead{Al$_2$O$_3$} &
	\colhead{C} &
	\colhead{Mg$_2$SiO$_4$}}
\startdata
{\bf 1}	&	0.039	&	 0.947	&	0.879	\\
{\bf 2}	&	0.184	&	 0.953 	&	0.902	\\
{\bf 3}	&	0.661	&	 0.964	&	0.938	\\
{\bf 4}	&	0.001	&	 0.714	&	0.369	\\
{\bf 5}	&	0.001	&	 0.714 	&	0.371	\\
{\bf 6}	&	0.118	&	 0.659	&	0.409	\\
{\bf 7}	&	0.001	&	 0.610	&	0.181 	\\
{\bf 8}	&	0.001	&	 0.597 	&	0.171	\\
{\bf 9}	&	0.000 	& 	 0.531 	&	0.137 	\\
{\bf 10}	&	0.002	&	 0.585	&	0.120	\\
{\bf 11}	&	0.002	&	 0.568	&	0.113	\\  
{\bf 12}	&	0.001	&	 0.442	&	0.067	\\ 
\vspace{-2.25mm}
\\
\tableline
\vspace{-2mm}
\\
{\bf \ Simulation}	&	MgSiO$_3$	&	SiO$_2$	&	MgO		\vspace{0.75mm}
\\
\tableline
\vspace{-1.5mm} 
\\
{\bf 1}	&	0.413	&	 0.910	&	0.827	\\
{\bf 2}	&	0.518	&	 0.931	&	0.857	\\
{\bf 3}	&	0.789	&	 0.951	&	0.922	\\
{\bf 4}	&	0.038	&	 0.411	&	0.347	\\
{\bf 5}	&	0.039	&	 0.413	&	0.351	\\
{\bf 6}	&	0.162	&	 0.441	&	0.399	\\
{\bf 7}	&	0.014 	&	 0.203	&	0.184	\\
{\bf 8}	&	0.014	&	 0.192	&	0.177	\\
{\bf 9}	&	0.007	&	 0.156	&	0.152 	\\
{\bf 10}	&	0.018	&	 0.128	&	0.132	\\
{\bf 11}	&	0.016	&	 0.121	&	0.126	\\  
{\bf 12}	&	0.008	&	 0.072	&	0.084	\\
\vspace{-2.25mm}
\\
\tableline
\vspace{-2mm}
\\
{\bf \ Simulation}	&	 Si 	&	 FeS 		&	 Fe 	\vspace{0.75mm}
\\
\tableline
\vspace{-1.5mm} 
\\
{\bf 1}	&	0.998	&	0.643	&	 1.000	\\
{\bf 2}	&	0.998	&	0.708	&	 1.000	\\
{\bf 3}	&	0.993	&	0.816 	&	 0.989	\\
{\bf 4}	&	0.964	&	0.165	&	 0.959	\\
{\bf 5}	&	0.960	&	0.145	&	 0.940	\\
{\bf 6}	&	0.815	&	0.173 	&	 0.693	\\
{\bf 7}	&	0.849	&	0.055	&	 0.801	\\
{\bf 8}	&	0.822 	&	0.044 	&	 0.752 	\\
{\bf 9}	&	0.660	&	0.007	&	 0.465	\\
{\bf 10}	&	0.726 	&	0.031	&	 0.603	\\
{\bf 11}	&	0.688	&	0.025	&	 0.547	\\  
{\bf 12}	&	0.446	&	0.005	&	 0.235
\enddata
\tablecomments{For simulation parameters, refer to Table \ref{sim_table}}
\end{deluxetable}

In studying the final (surviving) dust masses for each grain species, we find that the values are widely scattered for the simulations with the slower shock velocities.  The higher velocity simulations show dust survival of less than 20\% for six of the nine species, with C, Fe, and Si grains being the three species with the least destruction.  Although Fe grains had a higher survival rate than other dust species, the amount of Fe dust destruction was considerably higher than any of the simulations in our previous work.  In the most extreme case, only 24\% of the Fe dust mass survived.  If we look at dust survival for two other commonly studied grain species for the same extreme case, 44\% and 7\% of the initial dust mass remains for carbonaceous (C) and silicate (SiO$\subscript{2}$) grains, respectively.

For example, consider the differences in dust survival across grain species.  The difference, $\Delta \subscript{survival}$, between most destroyed grains and least destroyed grains ranges from as high as 96\% in Simulations 4 and 5 (Al$\subscript{2}$O$\subscript{3}$ compared to Si) to as low as 45\% in Simulation 12 (again, Al$\subscript{2}$O$\subscript{3}$ compared to Si).  This variation stems primarily from the initial distributions in grain radius.  A high degree of destruction is observed in the species with a majority of their mass locked up in small grains.  Destruction is reduced significantly for those species with appreciable mass in large grains.  As noted in Paper 1, grains with initial radii less than 0.1~$\mu$m are easily obliterated, while larger grains take a considerable duration to be substantially sputtered.  If the actual size distributions of dust grains in SNRs were greatly disparate from the distributions assumed here, the final dust mass survival fractions might change considerably.

\subsection{Density-Temperature Phase-Space}
\begin{figure*}[htp]
\centering
	\includegraphics[width=0.9\textwidth]{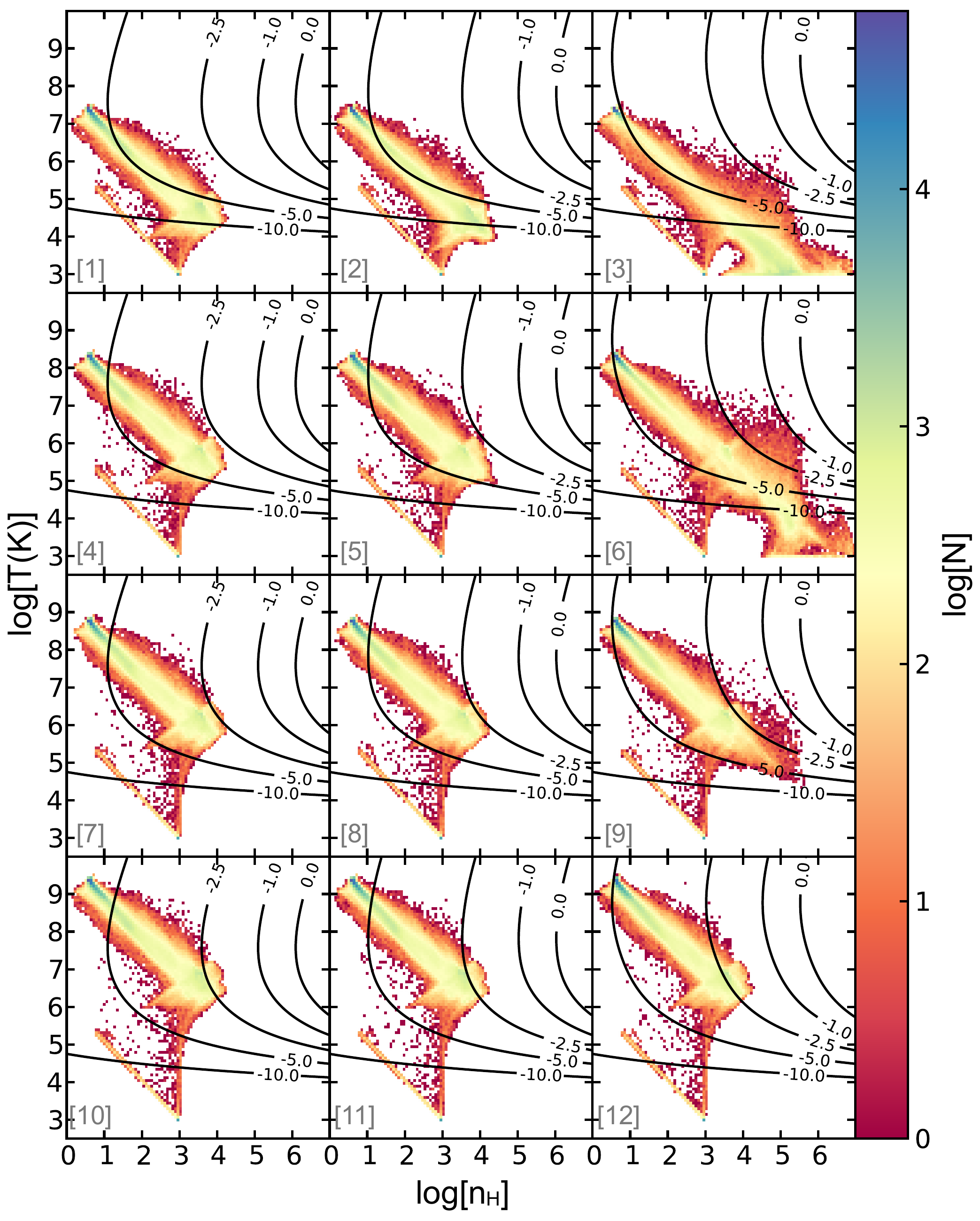}
	\caption{Phase plots of hydrogen number density ($n\subscript{H}$ in cm$^{-3}$) vs. temperature ($T$ in K).  The color scale represents the total number of times steps (N) spent by all dust particles in the simulation.  The contour lines show the dust destruction rate, log($da/dt$), in units of $\mu$m~yr$^{-1}$ for C grains. Each panel is one of the 12 simulations, labeled by the number in brackets [Table \ref{sim_table}].}\label{phaseplots}
\end{figure*}

In an attempt to understand the differences in the physical environments between the twelve simulations, we present phase-space diagrams of density and temperature to illustrate where the dust particles spend their time (Figure \ref{phaseplots}).  We also over-plot erosion rate contours for C grains as a function of density and temperature to determine where in phase-space the majority of the dust destruction occurs.  The most apparent difference between simulations is the rise in gas temperatures with increasing shock velocity, a direct result of shock-heating.  The increase in temperature pushes the gas into a regime of higher grain erosion, especially when the gas is still compressed to relatively high densities.  We also note that the large number of time steps at high temperature and low-density seen across all simulations is a result of the simulation end-state, at which point the cloud has been shock-heated and shredded.

In reviewing some of the subtle differences between simulations, we identify features that help to explain the differences between final dust masses.  First, we believe the high number of time steps spent by particles at high densities but very low temperatures in Simulation 3 ($v\subscript{shock} = 10^{3}$~km~s$^{-1}$, $Z = 100~Z\subscript{\odot}$) is likely the dominating factor in the decreased dust destruction when compared to the lower metallicity simulations (1 and 2). In this scenario, the temperatures of the gas drop low enough from the radiative cooling to slow grain erosion.  Although Simulation 6 ($v\subscript{shock} = 3 \times 10^{3}$~km~s$^{-1}$, $Z = 100~Z\subscript{\odot}$) shows a similar population of high-density, low-temperature dust particle environments, it is not sufficient to produce the same consistent decrease in dust destruction, owing primarily to the overall increase in gas temperature.  The pile up of dust particles at $T = 1000$~K, seen in Simulations 3 and 6, is a direct result of the temperature floor described in Section \ref{modifications}.  For $v\subscript{shock} = 5 \times 10^{3}$~km~s$^{-1}$, there is a distinct feature present in Simulation 9 with $Z = 100~Z\subscript{\odot}$ that separates it from the lower metallicity cases.  As a result of the enhanced metal cooling, the gas is able to cool and condense beyond a hydrogen number density of $n\subscript{H} = 10^{4}$ cm$^{-3}$, but it remains within a temperature regime to allow for appreciable grain erosion, explaining the increased dust destruction observed in the dust mass evolution.

In the case of the highest velocity shock, the intense heating appears to overwhelm the effects of radiative cooling, even for the highest metallicity simulations, as the phase plots for Simulations 10-12 appear nearly identical in the qualitative sense.  Given the similarity between the physical environments experienced by the dust particles, it would appear that in this regime the dominant cause for increased dust destruction at high metallicity is the increase in the erosion rates.  This is not an unreasonable explanation, since the dust particles spend the majority of their time above 10$^{6}$~K, the point at which the dust destruction rates begin to vary significantly for higher metal abundances.

\section{Summary and Discussion}\label{summary}
The above results can be summarized as follows:
\begin{itemize}
	\item{The degree of destruction varies widely across the explored parameter space. For the least destructive cases (low shock velocity, high gas metallicity), we see near-negligible grain destruction for some species (Fe and Si) and the highest survival rates for all other species. In the maximally destructive cases (high shock velocity, high gas metallicity), we find complete to near-complete destruction for multiple grain species (Al$_2$O$_3$, FeS, and MgSiO$_3$).}    
	\item{The relative velocity between the reverse shock and the cloud of ejecta material is a dominant factor in determining the survival rate of dust grains; an increase in velocity correlates directly with an increase in dust destruction.  The most extreme example is for silicate (SiO$\subscript{2}$) grains, where an additional 88\% of the initial dust mass is destroyed between the simulations with $v\subscript{shock} =  10^{3}$~km~s$^{-1}$ and those with $v\subscript{shock} =  10^{4}$~km~s$^{-1}$.}
	\item{There appears to be a threshold shock velocity between 1000 and 3000 km s$^{-1}$ at which the amount of dust destruction at early times ($t < t\subscript{cc}$) is increased considerably.}
	\item{In the physically expected, highly metal-enriched simulations, we find ranges of dust mass survival fraction of 44-96\%, 7-95\%, and 24-99\% for the often studied grain species of C, SiO$\subscript{2}$, and Fe, respectively.}
	\item{As in Paper I, the initial distribution in grain radius greatly influences the overall survival of a given species, with grains with initial radii below 0.1~$\mu$m being destroyed in most simulations.}
	\item{For the high-metallicity ($Z = 100~Z\subscript{\odot}$) simulations, the balance between the increased radiative cooling and increased sputtering yield depends strongly on the speed of the shock.  When the shock velocity is below 3000~km~s$^{-1}$, the increased cooling lowers the temperature of the gas enough to suppress dust destruction.  At high velocities, the gas temperature is driven up to a high-erosion regime.}
\end{itemize}

In comparing this new suite of simulations to that of Paper I, the key differences are the higher metallicity cooling functions, the erosion rates for enhanced metal abundances for all elements up to iron, the exploration of higher shock velocities ($v\subscript{shock} = 10^{4}$~km~s$^{-1}$), and the suppressed cooling in the ambient medium surrounding the cloud.  As a consistency check, we find that the simulations with $Z = Z\subscript{\odot}$ for shock velocities of $v\subscript{shock} = 1000$~km~s$^{-1}$, 3000~km~$^{-1}$, and 5000~km~s$^{-1}$ produce very similar dust mass survival fractions, indicating that these changes do not invalidate our previous results.  In fact, the degree of dust survival does not begin to deviate significantly from Paper 1 until the metallicity is increased to $Z = 100~Z\subscript{\odot}$, as expected.

We also find that our dust survival fractions remain in agreement with observations of IR emission in SNRs.  Specifically, the continued survival of Si grains matches the high abundances found in Cas A by \cite{Rho:2008qf}, while complete destruction of Al$_2$O$_3$ grains and survival of amorphous carbon grains match observations of SNR 1E0102-7219 by \cite{Sandstrom:2009fk}.  

Ideally, we would have liked to explore a metallicity regime in which the gas in the ejecta cloud is saturated with metals, effectively mimicking clouds of pure oxygen, silicon, sulfur, argon, or calcium.  Using oxygen as an example, an abundance ratio of order one would require a metallicity of $Z \sim 1000~Z\subscript{\odot}$ since the solar abundance ratio is $n\subscript{O}/n\subscript{H} \sim 5\times 10^{-4}$.  Unfortunately, due to the extremely short cooling times, this proved computationally prohibitive.  The simplest solution is to limit the simulation time-step by the cooling time and allow the simulation to run for a longer physical time, but it is also the most computationally expensive.  In order to probe the highest metallicity regimes, a more resource-efficient method would need to be implemented without sacrificing the validity of the simulation.  Such a solution is left to future work.

Currently, we still operate in the limit of strictly thermal sputtering, since our dust grains are directly coupled to the flow of the gas.  Because large grains would require a greater transfer of momentum from the gas before becoming completely entrained in the flow, it is possible that the dominant mode of sputtering for such grains would be in the non-thermal limit, as high velocity gas washes over them.  In order to decouple the dust particles from the motions of the gas, we would have to track separated grain populations of varying radii.  The number of radius bins required to accurately track the entire dust mass for all nine grain species is uncertain, but it could be computationally significant.  As mentioned in Paper I, it may not be unreasonable for the dust particles to be coupled to the flow if the grains are charged and magnetic fields permeate the ejecta material.  However, magnetized ejecta clouds would require an investigation of the effects of betatron acceleration on charged grains \citep{Shull:1977lr, Shull:1978fk}.

Finally, we investigated the possibility of transitioning from post-processing of our dust particle histories to compute erosion to an ``on-the-fly" method that would sputter the grains as the simulation ran.  This would allow metals that are sputtered off the grains to feed back into the ambient medium and increase the metallicity.  However, if we assume that the gas within the ejecta cloud is already heavily metal-enriched, the amount of metals released into the gas in this way would not alter the metallicity enough to make a substantial difference.  As we have shown, it takes a large change in the metal abundance to create an significant difference in the amount of dust that survives the shock-cloud interaction.

\acknowledgements
We thank the anonymous referee for comments that helped to clarify and strengthen this work.  We also thank Robert Fesen for useful discussion and comments.  The computing resources needed for this work were provided by NASA HEC Allocation SMD-10-2612 on the supercomputer, {\it Pleiades}, at the NASA Advanced Supercomputing Facility.  This project was funded by NSF Theory Grant AST07-07474, STScI Archive Theory Grant AR 11774.01-A, and the NSF Graduate Research Fellowship Program (D.W.S.).

\bibliography{references}

\end{document}